\documentclass[a4paper, conference]{IEEEtran}
\IEEEoverridecommandlockouts
\usepackage{cite}
\usepackage{amsmath,amssymb,amsfonts}
\usepackage{algorithmic}
\usepackage{graphicx}
\usepackage{textcomp}
\usepackage{xcolor}
\usepackage{mathtools}
\usepackage{adjustbox}
\usepackage{float}
\usepackage{tikz}
\usetikzlibrary{positioning}
\newcommand{\comment}[1]{}
\def\BibTeX{{\rm B\kern-.05em{\sc i\kern-.025em b}\kern-.08em
    T\kern-.1667em\lower.7ex\hbox{E}\kern-.125emX}}
    
\DeclarePairedDelimiterX{\infdivx}[2]{(}{)}{%
  #1\;\delimsize\|\;#2%
}
\newcommand{\infdiv}{KL\infdivx}

\usepackage{booktabs} 
\newcommand{\ra}[1]{\renewcommand{\arraystretch}{#1}}
\usepackage{siunitx}

\begin{document}

\title{More for Less: Non-Intrusive Speech Quality Assessment with Limited Annotations\\
\thanks{This  publication  has  emanated  from  research  conducted  with the financial support of Science Foundation Ireland (SFI) under Grant Number \mbox{17/RC-PhD/3483} and 17/RC/2289\_P2. The work of EB was supported by RAEng Research Fellowship RF/128 and a Turing Fellowship.}
}

\author{\IEEEauthorblockN{Alessandro Ragano$^{1,2}$, Emmanouil Benetos$^{3,4}$, and Andrew Hines$^{1,2}$}
\IEEEauthorblockA{\textit{$^{1}$ School of Computer Science, University College Dublin, Ireland \ \ \ $^2$ Insight Centre for Data Analytics, Ireland}
\\ \textit{$^3$ School of EECS, Queen Mary University of London, UK \ \ \ $^4$ The Alan Turing Institute, UK}\\
alessandro.ragano@ucdconnect.ie, emmanouil.benetos@qmul.ac.uk, andrew.hines@ucd.ie
}
}
\IEEEoverridecommandlockouts
\IEEEpubid{\makebox[\columnwidth]{978-1-6654-3589-5/21/\$31.00 \copyright 2021 IEEE \hfill} \hspace{\columnsep}\makebox[\columnwidth]{ }}
\maketitle

\begin{abstract}
Non-intrusive speech quality assessment is a crucial operation in multimedia applications. The scarcity of annotated data and the lack of a reference signal represent some of the main challenges for designing efficient quality assessment metrics. In this paper, we propose two \mbox{multi-task} models to tackle the problems above. In the first model, we first learn a feature representation with a degradation classifier on a large dataset. Then we perform MOS prediction and degradation classification simultaneously on a small dataset annotated with MOS. In the second approach, the initial stage consists of learning features with a deep \mbox{clustering-based} unsupervised feature representation on the large dataset. Next, we perform MOS prediction and cluster label classification simultaneously on a small dataset. The results show that the deep \mbox{clustering-based} model outperforms the degradation \mbox{classifier-based} model and  the 3 baselines (autoencoder features, P.563, and \mbox{SRMRnorm}) on \mbox{TCD-VoIP}. This paper indicates that \mbox{multi-task} learning combined with feature representations from unlabelled data is a promising approach to deal with the lack of large MOS annotated datasets.
\end{abstract}


\begin{tikzpicture}[overlay, remember picture]
\path (current page.north) node (anchor) {};
\node [below=of anchor] {%
2021 Thirteenth International Conference on Quality of Multimedia Experience (QoMEX)};
\end{tikzpicture}

\section{Introduction}
Speech quality assessment is fundamental to improve users' quality of experience (QoE) of multimedia communication systems. Perceived speech quality is affected by several degradations caused by many factors including audio codecs, network conditions, speech enhancement and background noise.  
The most accurate way to assess speech quality is through subjective listening tests. For instance, in the ITU standard P.800~\cite{ITUT1996}, participants judge speech quality on a \mbox{5-point} scale. Next, the sound quality of a stimulus is measured with the mean opinion scores (MOS) computed from several listeners. 

Despite their reliability, subjective listening tests are not always convenient given that they require a (1) substantial number of participants; (2) they cannot be used in \mbox{real-time} applications; (3) they are not suitable when large sound collections have to be evaluated; (4) they can be \mbox{time-consuming} and expensive. 

Objective quality metrics are a reliable replacement of the subjective judgement in the conditions above.
Objective quality metrics can be divided into \emph{\mbox{full-reference} metrics}, where a reference signal is available, and \emph{\mbox{non-intrusive} metrics} where quality is estimated through the noisy signal only. \mbox{Non-intrusive} objective metrics are preferred in scenarios where the reference signal does not exist such as real-time applications and \mbox{real-world} recordings. Traditional \mbox{non-intrusive} methods include the ITU standard P.563~\cite{malfait2006p} and SRMRnorm~\cite{santos2014improved}. A more recent approach to predict speech quality is to learn a mapping between noisy audio signals and MOS in a supervised learning fashion. The main drawback of using supervised learning is that a considerable amount of annotated data is required. Annotating large datasets is a general problem in machine learning, especially in multimedia quality assessment where multiple recruiters are needed to annotate only one stimulus. 
Although annotations obtained through crowdsourcing can be as valid as \mbox{lab-based} quality measures~\cite{naderi2020towards, hossfeld2013best}, annotating data is still a costly and \mbox{time-consuming} operation. 

To overcome the scarcity of annotated data, we propose unsupervised and supervised feature learning combined with \mbox{multi-task} learning. We introduce two \mbox{multi-task} learning techniques. 
In the first approach, we optimize both degradation classification and quality prediction simultaneously. Given the lack of large MOS annotated datasets, we propose to initialize the \mbox{multi-task} model by using the weights learned from a degradation classifier trained on a large dataset. Unlike quality prediction, classifying degradations can be done using large synthetic datasets where we apply various degradations. In the second approach, we propose a \mbox{semi-supervised} \mbox{multi-task} feature learning model without using the degradation labels. We first learn an unsupervised feature representation using deep convolutional embedded clustering (DCEC)~\cite{guo2017deep} on a large dataset where the MOS annotations are not given. Next, we cluster a small MOS annotated dataset using the feature representation and we use the cluster assignments as labels for the \mbox{multi-task} learning step. Our proposed approach can be especially useful in \mbox{real-world} scenarios where the knowledge of the degradation is not given and a large amount of \mbox{real-world} recordings is available~\cite{ragano2020development, ragano2019adapting}.

\section{Related Work and Motivation}
Recently, some \mbox{non-intrusive} metrics with deep learning techniques emerged. Only a few studies used large datasets annotated with MOS~\cite{avila2019non, cauchi2019non} while others relied on annotations created with \mbox{full-reference} metrics~\cite{fu2018quality, dong2019classification, catellier2020wawenets} or hybrid annotations~\cite{mittag2019non}.
Another group of \mbox{non-intrusive} metrics is closer to our approach, in the sense that they rely on different tasks to improve quality prediction. 
Ooster et al.~\cite{Ooster2018} used an automatic speech recogniser, assuming that phoneme posterior probabilities from a neural network degrade in presence of factors that affect speech quality. 
\mbox{Semi-Supervised} Speech Quality Assessment (SESQA)~\cite{serra2020sesqa} uses 5 complementary auxiliary tasks and 3 optimization criteria (MOS error, pairwise ranking, and score consistency). 
Soni et al.~\cite{soni2016novel} use a fully connected autoencoder to learn a feature representation from a large dataset. To the best of the authors' knowledge, the study of Soni et al. is the only one that learns an unsupervised feature representation for speech quality prediction. 

\mbox{Multi-task} learning~\cite{zhang2018overview} is based on training multiple tasks simultaneously. The motivation is that sharing weights between related tasks can improve all the tasks together. In our setting, MOS prediction is the main task and degradation classification or cluster assignment prediction is the auxiliary task. \mbox{Multi-task} learning improves generalisation and predictions of both tasks if the auxiliary task is related to the main task. 
Classifying degradations is a suitable auxiliary task for speech quality prediction because; (1) Perceived speech quality depends on the degradation~\cite{harte2015tcd}. The model has to learn how a clean speech signal is degraded, which is a concept associated with quality as well; (2) Classifying degradations is related to quality prediction but these tasks are not identical, which is desired~\cite{caruana1997multitask}. Indeed, two completely different degradations might be annotated with the same MOS.
Also, using degradation information together with quality prediction has been proposed in image quality assessment~\cite{kang2015simultaneous} and can be transferred to speech quality assessment similarly. In the second proposed approach, instead of relying on degradation labels, we generate cluster labels from unsupervised feature representations. The perceived quality of a speech signal is not only related to the degradation and more factors are involved. We assume that having labels that represent the similarity between the data points could be more meaningful than using only degradation labels in a \mbox{multi-task} learning scenario. These annotations are generated using only unlabelled data through a deep clustering~\cite{guo2017deep} technique that here we propose as a feature learning step for speech quality prediction.
It must be noted that different clustering techniques for unsupervised learning of features have been already employed in computer vision~\cite{caron2018deep, coates2012learning}

\section{Method}
In this section, we describe two methods for non-intrusive speech quality assessment based on Multi-Task Learning (MTL) and Semi-Supervised Multi-Task Learning (SEMTL).

\begin{figure}[t]
    \includegraphics[width=\columnwidth]{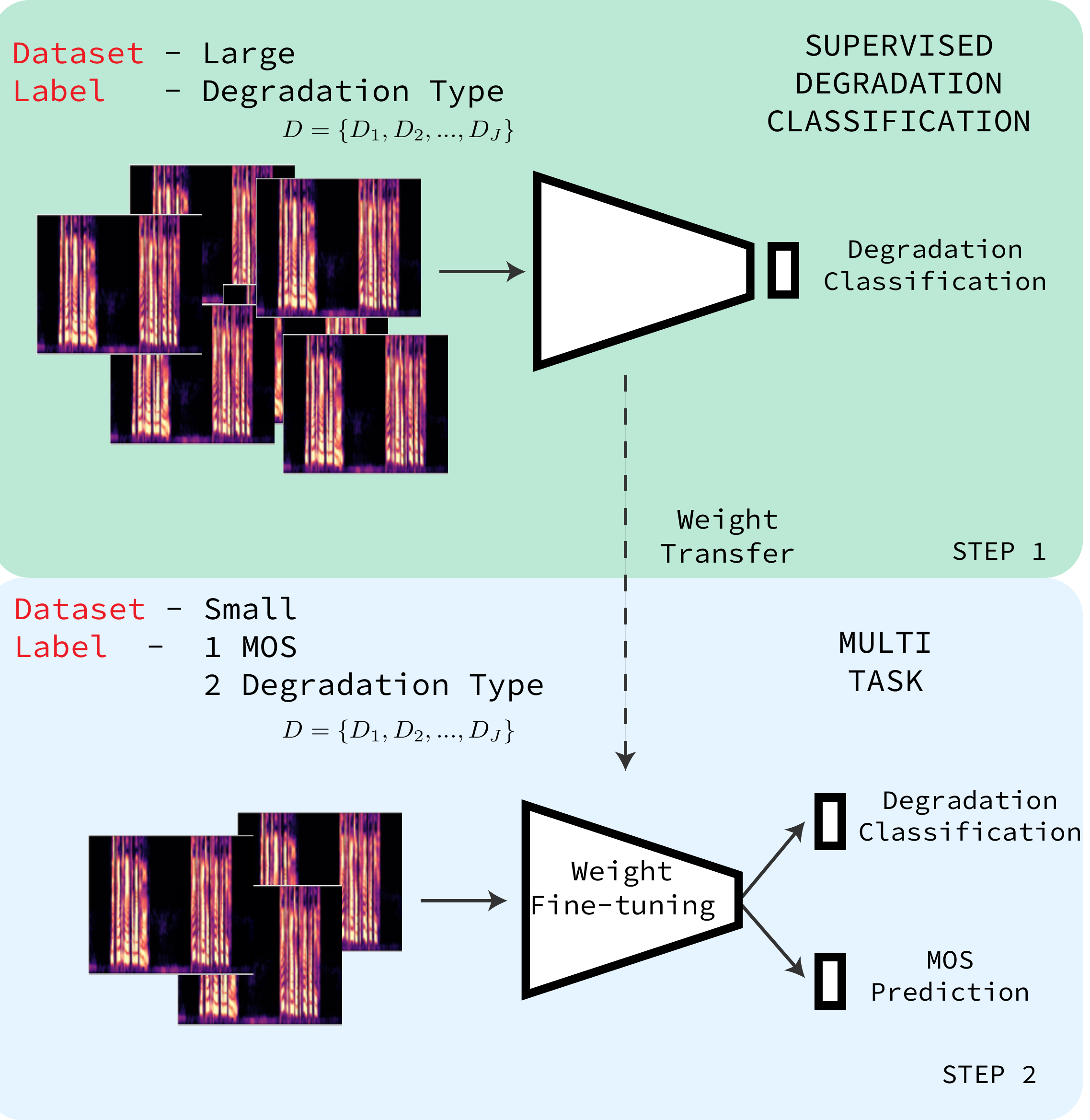}
    \caption{MTL-based model. STEP 1 consists of training a supervised degradation classifier on a large dataset. STEP 2 is a \mbox{multi-task} network trained on a small MOS annotated dataset that simultaneously classifies degradations and predicts quality scores. The model in STEP 2 is initialized with the weights learned in STEP 1.}
    \label{fig:MTL}
\end{figure}

\subsection{Multi-Task Learning}
In the \mbox{MTL-based} model, we train a model that performs simultaneous learning of MOS scores and degradation type. To tackle the lack of large MOS annotated datasets, we use the weights learned by a degradation classifier that is trained on a different and larger dataset. The approach is shown in Fig.~\ref{fig:MTL}. In the first step, we minimize the \mbox{cross-entropy}
$\mathcal{L}_{ce}(D,  \hat{D})$, where $D$ represents the degradation class and $\hat{D}$ is the predicted degradation type. We use a large dataset to learn a rich feature representation so that we can reuse the weights.  
In the second step, we initialize the weights from step 1 and we minimize the \mbox{multi-task} loss
\begin{equation}
\mathcal{L}_{tot} = \mathcal{L}_{ce}(D, \hat{D}) + \mathcal{L}_{mse}(S, \hat{S})
\end{equation}
where $S$ is the annotated MOS, $\hat{S}$ is the predicted score, and $\mathcal{L}_{mse}$ is the mean squared error. 
This second step is carried out on the small dataset where MOS annotations are available.

\subsection{Semi-Supervised Multi-Task Learning}
\subsubsection*{Motivation}
In the \mbox{SEMTL-based} model, we study whether a \mbox{multi-task} approach can be designed without using \mbox{human-annotated} labels in either dataset. We first learn a feature representation on the large dataset using deep convolutional embedded clustering (DCEC)~\cite{xie2016unsupervised, guo2017deep}. The motivation behind DCEC is that it simultaneously learns a feature representation and clusters the data on top of the feature representation. We use the DCEC cluster assignments as cluster labels on the annotated MOS dataset to perform \mbox{multi-task} learning without using the degradation information. The weights of the \mbox{multi-task} network are initialized with DCEC, similarly to the MTL approach above with the degradation classifier.  
We believe that using cluster labels might be beneficial for two reasons. First, the cluster assignments could represent concepts that are more complex than a degradation label. A speech signal is characterised by many factors which include rhythm, pitch, timbre and linguistic content. These factors are poorly represented by a degradation label. Instead, the cluster labels might represent a \mbox{high-level} similarity between the data points and as a consequence classification of such cluster labels can be seen as a useful auxiliary task. Secondly, when transferring the weights from DCEC we might deteriorate the feature representation due to the optimization of the weights for the target task i.e., MOS prediction. Therefore, when doing \mbox{multi-task} learning with the output of DCEC we help the network to retain existing knowledge of the learnt representation from the large dataset.

\begin{figure}[t]
    \includegraphics[width=\linewidth]{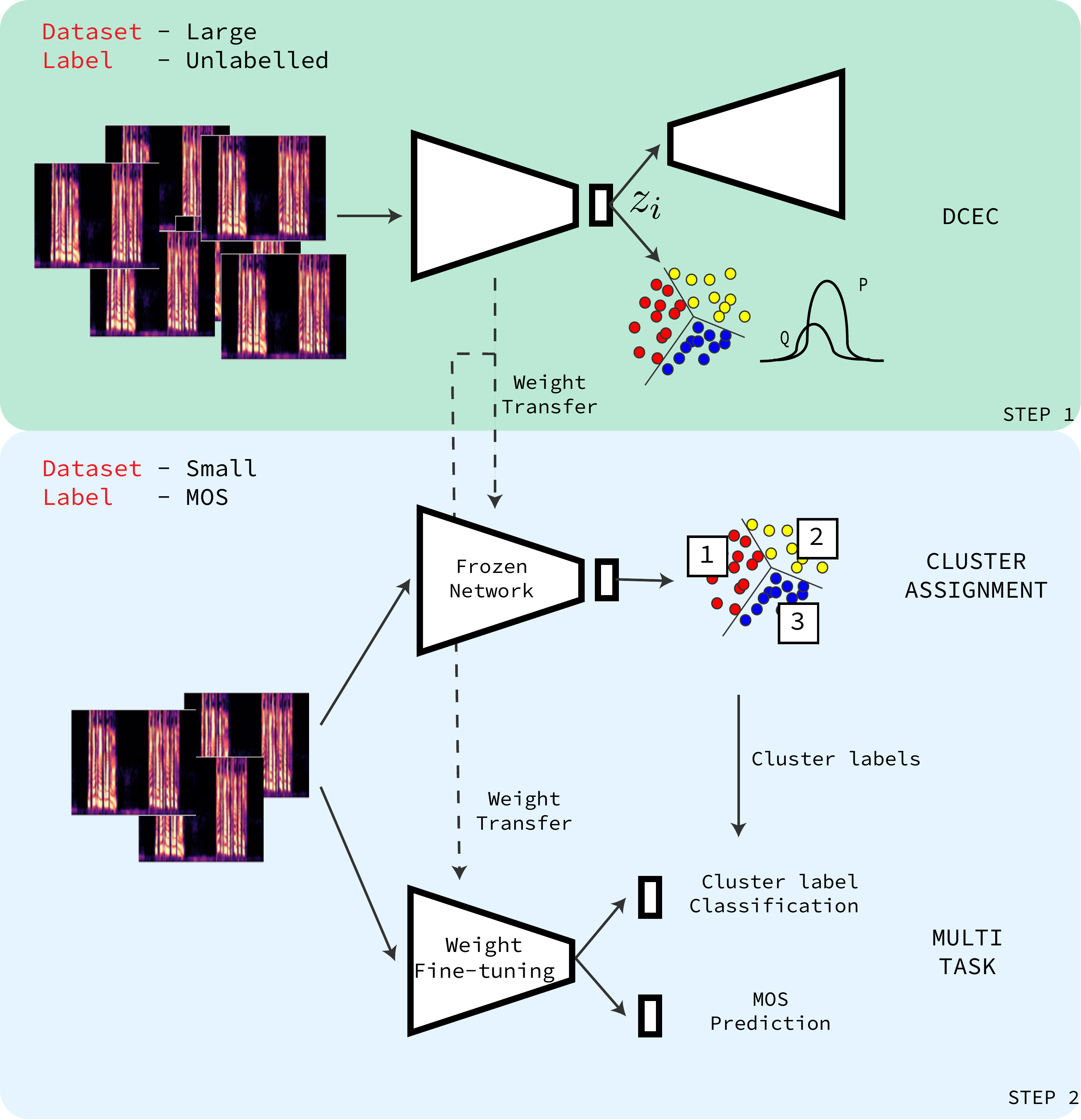}
    \caption{\mbox{SEMTL-based} model. STEP 1 consists of performing DCEC on the large set. In STEP 2 we first extract the cluster assignments on the small dataset using DCEC, which has been previously trained in STEP 1. Then, the \mbox{multi-task} network is trained to classify the cluster assignments and to predict MOS simultaneously. The model in STEP 2 is initialized with the weights learned in STEP 1.}
    \label{fig:semtl}
\end{figure}

\subsubsection*{SEMTL description}
The proposed \mbox{SEMTL-based} model is represented in Fig.~\ref{fig:semtl}.
In the first step, we use DCEC to cluster the large dataset. In the second step, we first assign clusters to the data in the small set by freezing the trained DCEC network. 
After annotating the dataset with the cluster assignments, we perform \mbox{multi-task} learning. The two tasks consist of cluster label classification and MOS prediction as follows:
\begin{equation}
\mathcal{L}_{tot} = \mathcal{L}_{ce}(Y^{(cl)}, \hat{Y}^{(cl)}) + \mathcal{L}_{mse}(S, \hat{S}) 
\end{equation}
where $Y^{cl}$ and $\hat{Y^{cl}}$ represent respectively the cluster labels and their prediction, $S$ is the MOS, $\mathcal{L}_{ce}$ is the \mbox{cross-entropy} and $\mathcal{L}_{mse}$ is the mean squared error. The \mbox{multi-task network} is initialized with the DCEC weights so that we use a learnt feature representation.

\subsubsection*{DCEC explained}
DCEC consists of a convolutional autoencoder and a clustering layer that is attached to the embedded layer of the autoencoder. 
The embedded features $z_{i}$ of the $i^{th}$ audio clip, are mapped by the clustering layer into a soft label using the Student's $t$-distribution: \begin{equation} q_{ij} = \frac{(1 + \| z_i - u_j \|^2/\alpha)^{-\frac{\alpha+1}{2}}}{\ \sum_{j'}^{}     (1 + \| z_i - u_{j'} \|^2/\alpha)^{-\frac{\alpha+1}{2}}} 
\end{equation}
where $q_{ij}$ is interpreted as the probability to assign the embedded point $z_i$ to a cluster $u_j$. The parameter $\alpha$ is set to $1$ in all the experiments as done in~\cite{guo2017deep}. The number of clusters $J$ is arbitrarily chosen and cluster centres are initialized by training \mbox{K-means} on embedded features of the convolutional autoencoder. DCEC is based on minimizing two loss functions as follows:
\begin{equation} 
\mathcal{L} = \mathcal{L}_{r} + \gamma \mathcal{L}_{c} 
\end{equation}
where $\mathcal{L}_{r} = |x - x'|^{2}_{2}$ is the cost function of the autoencoder with $x$ representing the input data. The second term is a clustering loss defined as $\mathcal{L}_{c} = \infdiv{P}{Q}$
which is the \mbox{Kullback-Leibler} divergence  between the soft label assignments $q_{ij}$ and an auxiliary target distribution $p_{ij}$ defined as:
\begin{equation} p_{ij} = \frac{ q_{ij}^2/f_{ij}}{\sum_{j'}^{}(q_{ij'}^2/f_{ij'})}
\end{equation} where $f_{ij} = \sum_{i} q_{ij}$.
The auxiliary target distribution is based on giving importance to the most confident predictions and penalizing clusters with too many samples. As in~\cite{guo2017deep} we fix $\gamma$ to $0.1$. 

\subsubsection*{DCEC optimization}
We optimize DCEC with convolutional autoencoders as done in~\cite{guo2017deep}. We first train an autoencoder to initialize the DCEC network parameters and the cluster centres. Then we minimize the DCEC cost function which updates both cluster centres $u_{j}$ and network parameters. All the learnable parameters can be updated with backpropagation as shown in~\cite{guo2017deep, xie2016unsupervised}. The auxiliary target  distribution is updated every $T$ iterations to avoid instability.

\section{Experiments}
\subsection{Datasets}
\mbox{TCD-VoIP}~\cite{harte2015tcd} is used as the small MOS annotated dataset which consists of 384 recordings sampled at 48 kHz. This dataset size is typically considered as not sufficient for training an efficient deep learning model\footnote{In the results section we show the performance of a naive baseline model trained on this dataset only.}. We take the audio stimuli with the following degradations: chop, clip, echo and background noise, collecting around 38 minutes of data. 
The dataset consists of degraded stimuli created with clean speech taken from the TSP speech database~\cite{kabal2002tsp}. Speech sentences have a duration of $\approx$8 seconds and include 4 speakers (2 Male, 2 Female). \mbox{TCD-VoIP} includes several conditions for each degradation and there are 4 clips for each condition. 

The large annotated dataset is built with the same degradations of \mbox{TCD-VoIP} using different speakers and sentences from the TSP database. TSP includes 24 adult speakers. We discarded the 4 speakers used in \mbox{TCD-VoIP} and we use the remaining 20 speakers. 
In this way, we make sure that no speaker or sentence dependent biases are transferred from the model trained on the large dataset to the one trained on \mbox{TCD-VoIP}. 
For each degradation, we include more conditions than the ones present in the \mbox{TCD-VoIP} to improve generalization in our model. 
We generate 3805 recordings which are almost 8.5h of audio. The dataset is divided into 761 stimuli per class. In total, we use 5 classes: CHOP, CLIP, ECHO, NOISE, REFERENCE. The reference speech is included as distinguishing degraded speech from clean speech could be useful for quality prediction. 

\subsection{Experiment set-up}
Initial experiments showed that using 48~kHz sampling was not adding any benefit. Therefore, we downsampled the data to 16~kHz to reduce the input dimension. We transformed each raw audio waveform to log mel spectrograms using 64 mel bands and windows of $25$ ms with $10$ ms hop length. In both MTL and SEMTL we used 5 classes as we want to equally compare SEMTL with MTL. Therefore, we selected $5$ clusters for DCEC and $5$ classes for the auxiliary task in SEMTL (i.e., we classify 5 cluster labels in the auxiliary task). 
DCEC is trained in two steps. First, we trained an autoencoder for 200 epochs. Secondly, DCEC with the clustering loss is trained until the number of cluster assignments between two consecutive auxiliary target distribution updates is lower than a threshold. We set the convergence threshold to $0.1\%$ of the dataset size. We updated the auxiliary target distribution every $70$ batches. In~\cite{guo2017deep} the target distribution is updated every $140$ steps but our experiments showed instability. 
The supervised classifier is trained for 200 epochs as well. In all the experiments on the large dataset, we use a batch size of $64$ and we update the weights using Adam optimizer with a learning rate of $0.001$. 

\begin{figure*}[!t]
    \includegraphics[width=\linewidth]{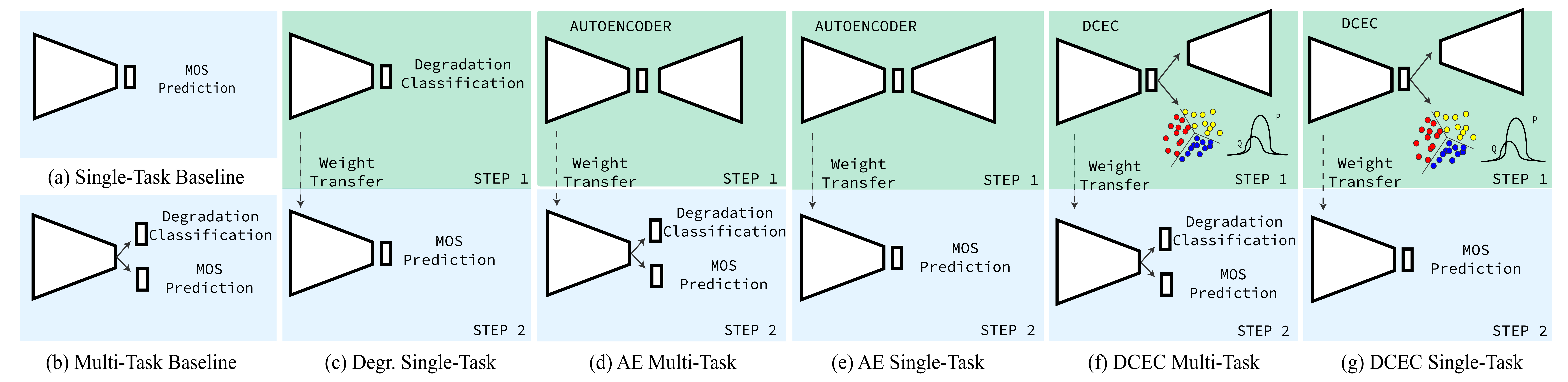}
    \caption{Additional models that we used in the experiments. Step 1 is trained with the large dataset while step 2 with \mbox{TCD-VoIP}. In step 2 we do \mbox{fine-tuning} in each model. The two baselines (a) and (b) are \mbox{end-to-end}.}
    \label{fig:blocks}
\end{figure*}

The quality prediction networks are trained and tested in a \mbox{speaker-independent} scenario. We partition the \mbox{TCD-VoIP} dataset using \mbox{4-fold} \mbox{cross-validation} so that 3 speakers are in the training set and the remaining speaker is in the test set. This is done to avoid that the same sentence appears in both training and test set. The \mbox{multi-task} models after \mbox{pre-training} are optimized for $40$ epochs in each fold, using Adam with a learning rate of $0.00001$ and a batch size of $64$. 

\subsection{Architecture}
We use the same convolutional architecture (ConvNet) in every model and we attach different fully connected layers depending on the task. Fixing the same ConvNet is required so that we can transfer the weights and we can fairly compare the different feature representations. The ConvNet consists of 4 layers $L_{32}^{5} \rightarrow{} L_{64}^{5} \rightarrow{} L_{128}^{3} \rightarrow{} L_{256}^{3} \rightarrow{}$ where $L_{m}^{k}$ means a convolutional layer with $m$ kernels and $(k, k)$ kernel size. We used stride $2$ and ``same" padding in all the layers. This architecture represents the encoder in the autoencoder used in DCEC and the convolutional part in all of the other models. The structure of the decoder is the mirror of the encoder. In each layer, we use the ReLU activation function and batch normalization. For each task, we attach a fully connected network to the ConvNet as summarised in Table \ref{table:fcl}. 

\begin{table}[t]\centering
\ra{0.4}. 
\centering

\caption{Fully connected networks attached to the ConvNet based on the task type. When we classify degradations or cluster assignments we attach ``Classification". When we predict MOS scores we attach the network ``Regression". We use a shifted sigmoid $\sigma$ as the final activation function to map the MOS range from 1 to 5.
DCEC is composed by two fully connected layers, 10 neurons for the embedded layer and 5 neurons for the clustering layer. In each \mbox{multi-task network}, we attach both ``Classification" and ``Regression" to the same shared ConvNet.}
\begin{Huge}{}
\begin{adjustbox}{max width=0.50\textwidth,center}
\begin{tabular}{>{\small}l>{\small}l>{\small}c} 
\toprule
 \textbf{Task} & \textbf{Model} \\ 
 \midrule
 DCEC           & ConvNet $\rightarrow{}$ 
                $\textrm{FC}_{10}$ $\rightarrow{}$ $\textrm{FC}_{5}$\\  
 Classification & ConvNet $\rightarrow{}$ 
                $\textrm{FC}_{256}$ $\rightarrow{}$ ReLU $\rightarrow{}$ D(0.5) $\rightarrow{}$ $\textrm{FC}_{5}$ $\rightarrow{}$ Softmax\\ 
 Regression                        & ConvNet $\rightarrow{}$ 
 $\textrm{FC}_{256}$ $\rightarrow{}$ ReLU $\rightarrow{}$ D(0.5) $\rightarrow{}$ $\textrm{FC}_{1}$ $\rightarrow{}$ $1 + 4\sigma$\\
\bottomrule
\end{tabular}
\end{adjustbox}
\end{Huge}
\label{table:fcl}
\end{table}

\subsection{Results}
In our experiments, we want to compare MTL and SEMTL to each other. Also, we compare our proposed models with different combinations of feature representations with \mbox{single-task} or \mbox{multi-task} scenarios as shown in Fig.~\ref{fig:blocks}. We also explore the \mbox{multi-task} model using degradation classification as an auxiliary task after learning features with a convolutional autoencoder trained on the large dataset. The autoencoder is the same that we use to initialize DCEC. 

We test the predicted scores against MOS using \mbox{root-mean-square} error (RMSE), Pearson correlation coefficient (PCC) and Spearman's \mbox{rank-order} correlation coefficient (SRCC)~\cite{ITUT2020} as shown in Table \ref{table:results}. Statistics are calculated in each fold and the average is reported.  
Results show that every \mbox{multi-task} model combined with feature learning outperforms its respective \mbox{single-task} \mbox{pre-trained} counterpart as well as the two baselines P.563 and SRMRnorm\footnote{P.563 and SRMRnorm are computed from https://github.com/qin/p.563 and https://github.com/MuSAELab/SRMRToolbox on data downsampled to 8 kHz.}. 
The models that show the highest correlations with subjective ratings are SEMTL and MTL while the \mbox{single-task} models \mbox{pre-trained} with the autoencoder and the degradation classifier shows the highest RMSE.

\begin{table}[t]
\ra{0.4}. 
\centering
\caption{Performance evaluation with Root Mean Square Error (RMSE), Pearson Correlation Coefficient (PCC), and Spearman's Rank-Order Correlation Coefficient (SRCC).}
\begin{Huge}{}
\begin{adjustbox}{max width=0.5\textwidth,center}
\small
\begin{tabular}{>{\small}l|>{\small}c>{\small}c>{\small}c>{\small}c|>{\small}c|>{\small}c} 
 \textit{RMSE} & \textbf{CHOP} & \textbf{CLIP} & \textbf{ECHO} & \textbf{NOISE} & \textbf{ALL}  &\textbf{P.Supp23} \\ 
 \midrule
  \textbf{Single-Task Baseline} & $0.832$ & $0.825$ & $0.889$ & $0.573$ & $0.786$ & $0.621$ \\ 
   \textbf{Multi-Task Baseline} & $0.81$ & $0.779$ & $1.039$ & $0.667$ & $0.85$ & / \\
  \textbf{Degr. Single-Task} & $0.787$ & $0.74$ & $\mathbf{0.758}$ & $\mathbf{0.377}$ & $\mathbf{0.684}$ & $\mathbf{0.574}$ \\
   \textbf{AE Multi-Task} & $0.882$ & $0.711$ & $\mathbf{0.779}$ & $0.416$ & $0.709$ & / \\
  \textbf{AE Single-Task} & $0.85$ & $\mathbf{0.664}$ & $0.812$ & $0.423$ & $\mathbf{0.698}$ & $0.601$ \\
 \textbf{DCEC Multi-Task} & $0.765$ & $\mathbf{0.703}$ & $0.8$ & $0.447$ & $0.728$ & /\\  
 \textbf{DCEC Single-Task} & $\mathbf{0.734}$ & $0.742$ & $0.88$ & $0.525$ & $0.744$ & $0.591$ \\
 \textbf{MTL} & $\mathbf{0.755}$ & $0.799$ & $0.817$ & $0.412$ & $0.705$ & /\\
 \textbf{SEMTL} & $\mathbf{0.755}$  & $0.802$ & $0.814$ & $\mathbf{0.407}$ & $0.703$ & $\mathbf{0.586}$ \\ 
\midrule
 \textbf{P.563} & $0.801$  & $0.762$ & $1.04$ & $1.11$ & $0.966$ & $1.804$\\ 
  \textbf{SRMRnorm} & $1.01$  & $1.103$ & $1.061$ & $1.055$ & $1.055$ & $0.766$ \\

\bottomrule
\end{tabular}
\end{adjustbox}
\end{Huge}

\ra{0.4}
\vspace{1em}
\centering

\begin{Huge}{}
\begin{adjustbox}{max width=0.5\textwidth,center}
\small
\begin{tabular}{>{\small}l|>{\small}c>{\small}c>{\small}c>{\small}c|>{\small}c|>{\small}c} 
 \textit{PCC} & \textbf{CHOP} & \textbf{CLIP} & \textbf{ECHO} & \textbf{NOISE} & \textbf{ALL} & \textbf{P.Supp23}   \\ 
 \midrule
 \textbf{Single-Task Baseline} & $0.335$ & $0.532$ & $0.679$ & $0.868$ & $0.659$ & $0.551$ \\ 
 \textbf{Multi-Task Baseline} & $0.446$ & $0.688$ & $0.491$ & $0.882$ & $0.622$ & / \\
 \textbf{Degr. Single-Task} & $0.437$ & $\mathbf{0.810}$ & $\mathbf{0.821}$ & $\mathbf{0.941}$ & $0.770$ & $\mathbf{0.691}$ \\
 \textbf{AE Multi-Task} & $0.223$ & $0.758$ & $\mathbf{0.807}$ & $0.927$ & $0.748$  & / \\
 \textbf{AE Single-Task} & $0.361$ & $0.775$ & $0.719$ & $0.920$ & $0.746$  & $0.565$ \\
 \textbf{DCEC Multi-Task} & $0.493$ & $0.753$ & $0.775$ & $0.907$ & $0.726$ & /\\  
  \textbf{DCEC Single-Task} & $0.557$ & $0.685$ & $0.66$ & $0.867$ & $0.699$ & $0.600$ \\
 \textbf{MTL} & $\mathbf{0.588}$ & $0.739$ & $0.794$ & $0.932$ & $\mathbf{0.773}$ & /\\
  \textbf{SEMTL} & $\mathbf{0.580}$ & $0.736$ & $0.801$ & $\mathbf{0.935}$ & $\mathbf{0.776}$ & $\mathbf{0.608}$  \\ 
\midrule
 \textbf{P.563} & $0.492$  & $0.707$ & $0.551$ & $0.003$ & $0.357$ & $0.507$ \\ 
   \textbf{SRMRnorm} & $0.425$  & $\mathbf{0.8}$ & $0.512$ & $0.596$ & $0.511$ & $0.386$\\ 
\bottomrule
\end{tabular}
\end{adjustbox}
\end{Huge}

\ra{0.4}
\vspace{1em}
\centering

\begin{Huge}{}
\begin{adjustbox}{max width=0.5\textwidth,center}
\small
\begin{tabular}{>{\small}l|>{\small}c>{\small}c>{\small}c>{\small}c|>{\small}c|>{\small}c} 
 \textit{SRCC} & \textbf{CHOP} & \textbf{CLIP} & \textbf{ECHO} & \textbf{NOISE} & \textbf{ALL} & \textbf{P.Supp23}   \\ 
 \midrule
  \textbf{Single-Task Baseline} & $0.322$ & $0.476$ & $0.614$ & $0.807$ & $0.618$ & $0.574$\\ 
 \textbf{Multi-Task Baseline} & $0.461$ & $0.744$ & $0.534$ & $0.847$ & $0.598$ & / \\
 \textbf{Degr. Single-Task} & $0.423$ & $0.738$ & $\mathbf{0.82}$ & $\mathbf{0.885}$ & $0.729$ & $\mathbf{0.696}$\\
 \textbf{AE Multi-Task} & $0.193$ & $0.792$ & $0.755$ & $0.873$ & $0.689$ & / \\
  \textbf{AE Single-Task} & $0.33$ & $\mathbf{0.823}$ & $0.595$ & $0.86$ & $0.683$ & $0.565$ \\
 \textbf{DCEC Multi-Task} & $0.54$ & $\mathbf{0.805}$ & $0.729$ & $0.857$ & $0.674$ & / \\  
 \textbf{DCEC Single-Task} & $0.536$ & $0.695$ & $0.587$ & $0.817$ & $0.648$ & $\mathbf{0.633}$ \\
 \textbf{MTL} & $\mathbf{0.565}$ & $0.736$ & $0.790$ & $0.875$ & $\mathbf{0.754}$ & / \\
 \textbf{SEMTL} & $\mathbf{0.556}$ & $0.792$ & $\mathbf{0.791}$ & $\mathbf{0.884}$ & $\mathbf{0.76}$ & $0.632$ \\ 
\midrule
 \textbf{P.563} & $0.478$  & $0.69$ & $0.501$ & $0.013$ & $0.339$ & $0.491$\\ 
   \textbf{SRMRnorm} & $0.363$  & $0.753$ & $0.509$ & $0.544$ & $0.475$  & $0.390$ \\ 
\bottomrule
\end{tabular}
\end{adjustbox}
\vspace{-0.5em}
\end{Huge}
\label{table:results}
\end{table}

The same procedure described in step 2 of each proposed model was repeated using the P.Supp23 Experiment 1 database~\cite{ITUT1998}, which is a speech codec dataset including different languages (Japanese, French, and English). We test unseen degradations to evaluate the generalization capacity of the proposed approach. Therefore, we only transferred the weights from the model trained on the same large dataset and we \mbox{fine-tuned} on P.Supp23 Experiment 1 using cross-validation. We split the dataset by speakers so that 3 speakers, 1 per language, are in the test set and the remaining speakers in the training set. 
P.Supp23 Experiment 1 results (Table~\ref{table:results}) show that SEMTL has poor adaptation performance while the degradation classifier has the best generalization capacity. It should be noted that SEMTL assigned 3 cluster labels roughly with the following proportion 86\%, 8\%, and 6\%. Given that most of the clips belong to the same cluster, the multi-task step with cluster labels is not expected to be useful in this scenario which is aligned with the results. The results suggest that the proposed approach might not be appropriate for datasets not too varied in terms of degradations such as P.Supp23 Experiment 1 that only includes speech codecs.

\subsection{Cluster analysis}
\begin{figure*}
  \centering
  \begin{minipage}{0.49\linewidth}
    \centering
    \includegraphics[width=0.58\textwidth]{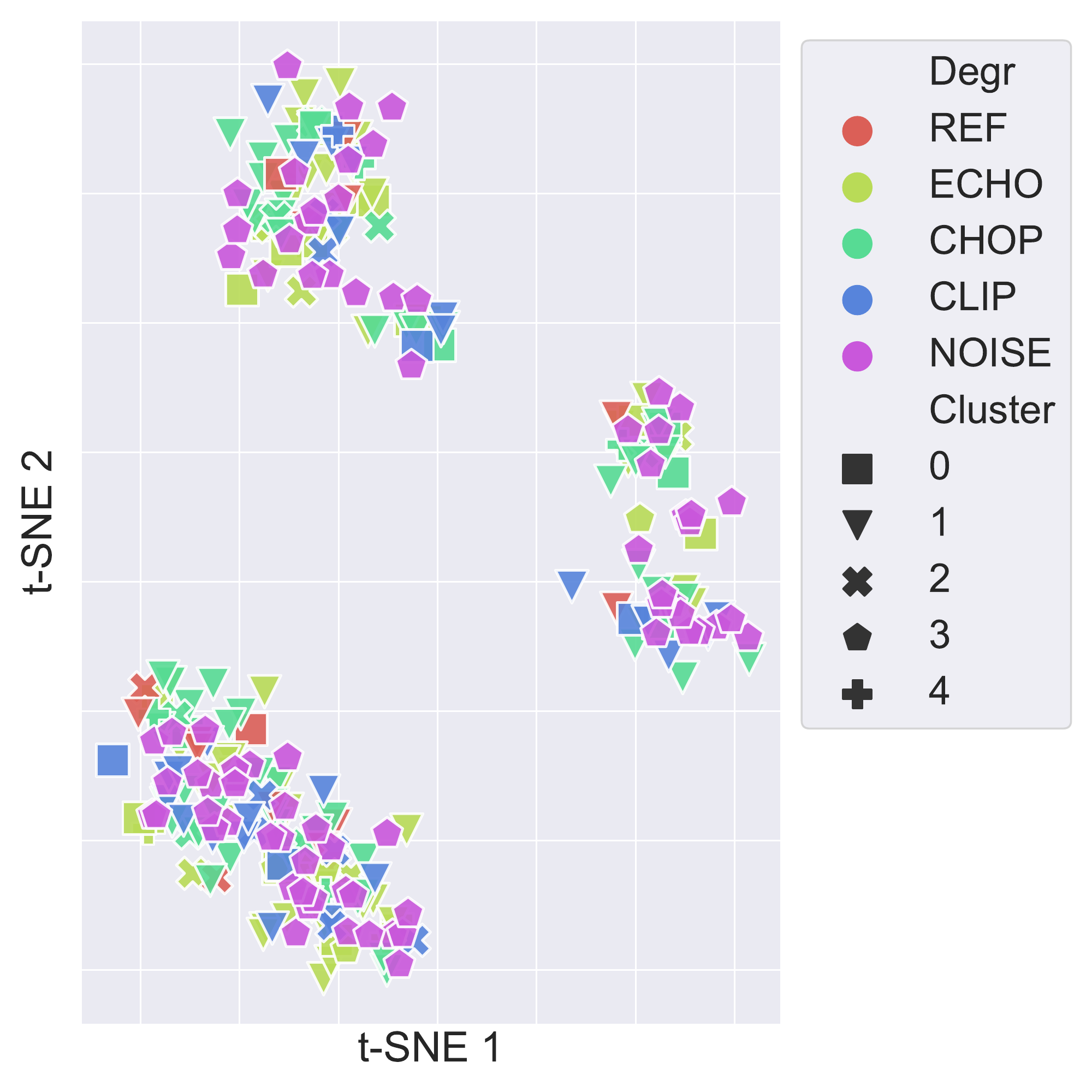}
    \phantom{\rule[1ex]{14em}{0.5pt}}
    \end{minipage}
  \begin{minipage}{0.49\linewidth}
    \centering
    \includegraphics[width=0.64\textwidth]{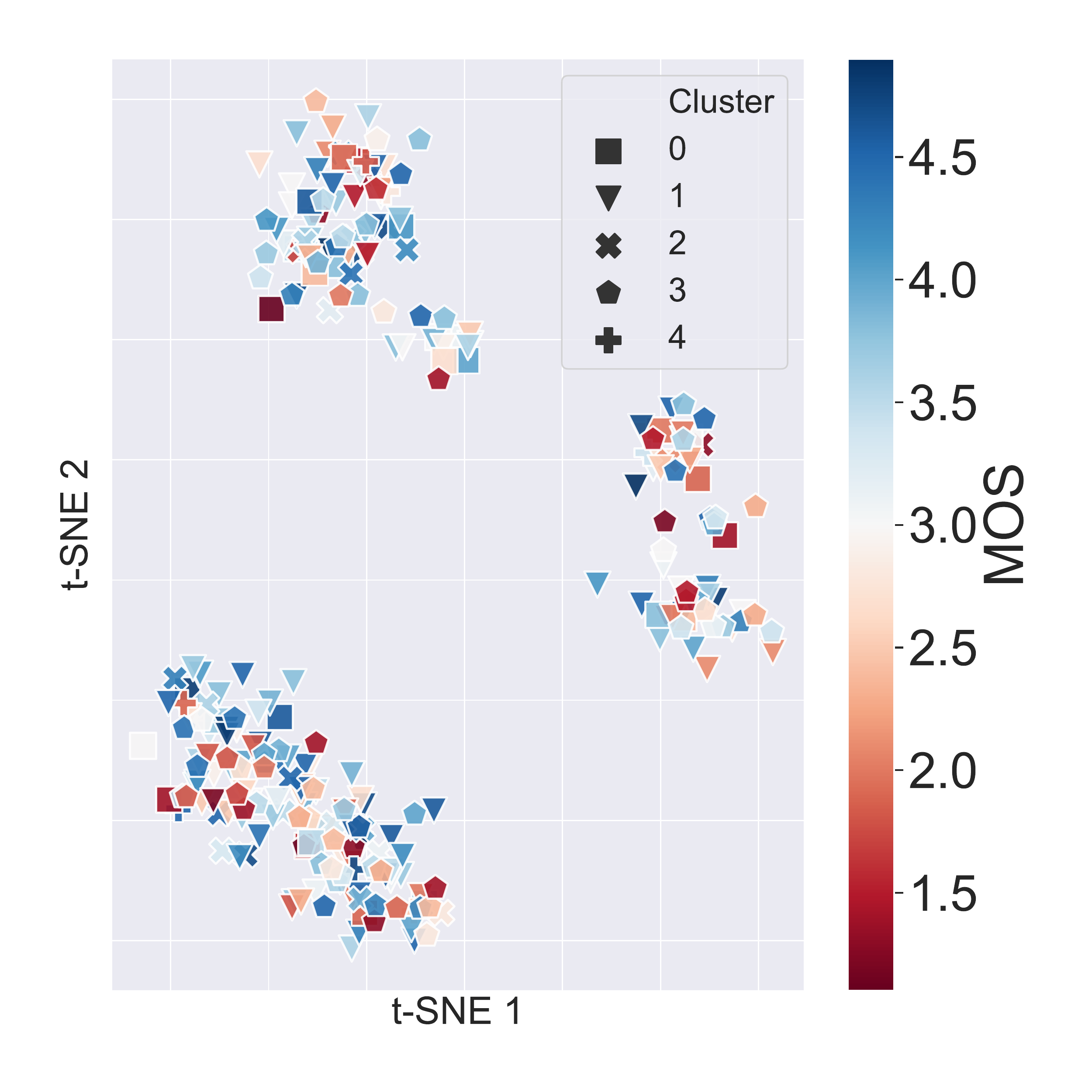}
    \phantom{\rule[1ex]{14em}{0.5pt}}
    \end{minipage}
\caption{Comparing DCEC clusters with degradations (left) and MOS (right) using \mbox{t-SNE}. "REF" data points represent the first condition of each degradation.}
\label{fig:dcec_clusters}
\vspace{-1em}
\end{figure*}

Fig.~\ref{fig:dcec_clusters} suggests that DCEC performs clustering according to a criterion that does not correspond to either degradations or MOS. Our results show that the simultaneous classification of the cluster assignments is beneficial for quality prediction which suggests that a \mbox{high-level} grouping of the data points might occur. 

\section{Discussion}
We have shown that \mbox{multi-task} learning combined with unsupervised and supervised feature learning is a promising approach to deal with limited annotations in speech quality assessment. In particular, the SEMTL-based model is fully unsupervised and might allow using uncurated data from \mbox{real-world} recordings which is generally advantageous. The results of this paper suggest that going towards a better feature representation from unlabelled data is a promising approach and that might be taken into consideration as opposed to collecting large annotated MOS datasets which is not always affordable. Although the proposed techniques show less promising performance in the case of different domains, unsupervised domain adaptation techniques~\cite{lordelo2020adversarial} could be explored in the future to mitigate the domain mismatch issues.

\section{Conclusions and Future Work}
In this paper, we have proposed two \mbox{multi-task} learning approaches combined with unsupervised feature learning for \mbox{non-intrusive} speech quality assessment. In the \mbox{MTL-based} approach, we use degradation classification either as an auxiliary task and for learning the initial weights. The \mbox{SEMTL-based} approach consists of classifying cluster labels generated from DCEC, a deep clustering technique that learns features and clusters the data simultaneously. We have shown that \mbox{multi-task} learning combined with unsupervised feature learning shows promising performance for \mbox{non-intrusive} speech quality assessment using a very small MOS annotated dataset. In particular, SEMTL does not need any auxiliary labels and achieves better performance than DCEC \mbox{multi-task} which uses degradation classification as an auxiliary task. 

In the future, we will evaluate this approach with a larger dataset. We will design a transfer learning approach where we take a dataset with degradations from different applications (e.g., speech enhancement, audio codecs etc.).
The experiments shown in this paper do not explore the number of clusters in DCEC. We believe that feature representation learned with DCEC might be sensitive to the number of clusters and that the optimal number of clusters could have not been found in this paper.
Finally, we are aware that we have used a basic \mbox{multi-task} approach (e.g., we have not found optimal weight loss). We will explore different \mbox{multi-task} techniques with both degradation types and cluster assignments. 

\bibliographystyle{unsrt}
\bibliography{bibliography}
\end{document}